**Non-Affine Displacements and the Non-Linear Response of a Strained Amorphous Solid**

Shibu Saw, Sneha Abraham and Peter Harrowell

School of Chemistry, University of Sydney, Sydney 2006 NSW, Australia

Abstract

We demonstrate that irreversible structural reorganization is *not* necessary for the observation of yield behaviour in an amorphous solid. While the majority of solids strained to their yield point do indeed undergo an irreversible reorganization, we find a significant fraction of solids exhibit yield via a reversible strain. We also demonstrate that large instantaneous strains in excess of the yield stress can result in complete stress relaxation, a result of the large non-affine motions driven by the applied strain. The empirical similarity of the dependence of the ratio of stress over strain on the non-affine mean squared displacement with that for the shear modulus obtained from quiescent liquid at non-zero temperature supports the proposition that rigidity depends on the size of the sampled configurational space only, and is insensitive as to how this space is sampled.

## 1. Introduction

The limit of rigidity of a material can be measured by the decrease of the shear modulus as a function of the magnitude of an applied strain. In crystalline materials, this loss of the modulus has been shown to result from the motion of defects, either pre-existing or generated under strain [1]. The success of this microscopic description of nonlinear mechanical response represents one of the cornerstones of materials science [2]. In 1979, Argon and co-workers [3] proposed an extension of the idea of localised defects to account for the nonlinear mechanical response of amorphous solids. In the absence of an explicit structural definition, these defect-like objects, referred to as shear transformation zones, were characterised as localised irreversible reorganization events and, as such, have been widely used to treat the mechanical response of metallic glasses [4]. Recently, a number of groups [5,6] have challenged the notion that the irreversible reorganization events in an amorphous solid under shear are localised. Simulation result of quasi-static shearing at zero temperature have reported that the plastic reorganization events are extended, spanning the length of the simulation cell [5,6]. If the microscopic mechanisms responsible for stress relaxation in amorphous solids are not localized, do they even need to be irreversible? In this paper we shall present evidence that *reversible* strains can account for the non-linear response of an amorphous solid under shear. We shall also demonstrate that instantaneous strains larger than the yield strain result in the relaxation of stress and argue that this relaxation is related to the magnitude of the non-affine motions in a manner quantitatively similar to that observed in quiescent materials at non-zero temperatures. This latter result allows us to establish a fundamental connection between the shear-induced loss of rigidity and that achieved in the quiescent material by heating.

## 2. Model

The model liquid used in this study is a 2D system of soft disks with a pair interaction potential, $\phi_{ij}(r) = \varepsilon \left( \dfrac{a_{ij}}{r} \right)^{12}$, between species i and j. We consider an equimolar binary mixture with $a_{11}$=1.0, $a_{22}$=1.4 and $a_{12}$=1.2 and all particle with unit mass, a model that has been extensively studied [7] in the context of the glass transition. The following reduced units are used: $\varepsilon/k_B$ for temperature T and times are reported in units of $\tau = \sqrt{ma_{11}^2/\varepsilon}$. Simulations were carried out under constant NVT conditions using LAMMPS [8] with a Nose-Hoover thermostats at a reduced density 0.7468 with a potential cut-off distance of $6.3a_{11}$. The system consisted of a total of N = 1024 particles. We have generated 51 distinct local minima of the potential energy (referred to here as an *inherent structure* [9]). In each case the liquid was cooled at a rate of 5 x $10^{-5}$ from T=0.60 to T=0.30 and the resulting configuration then subjected to a potential energy minimization to obtain the final inherent structure configuration. We note that increasing the rate of cooling or the temperature of the parent liquid will give rise to a decrease in the shear modulus and an associated increase in the non-affine displacements associated with applied strain [10]

## 3. Non-Linear Mechanical Response by Reversible Strains

We apply an affine shear strain of magnitude γ in a single step to an initial configuration (inherent structure) corresponding to a local potential energy minimum with the associated application of Lees-Edwards boundary conditions [11]. Following this affine strain, the potential energy of the strained configuration is then subjected to a conjugate gradient minimization, under the constraint of the strained boundary conditions and the resulting shear stress σ is calculated. The shear stress σ, averaged over the inherent structures, is plotted against strain in Fig. 1. The non-linear mechanical response is clearly evident in the deviation of <σ> from the initial linear dependence on strain. The shear stress exhibits a maximum at a yield strain $\gamma^* = 0.05$ followed by a steady reduction in the stress with further increase in γ.

This strain-induced stress relaxation is quite different from the mechanical behaviour found when a large shear strain is applied quasi-statically. In the quasi-static protocol [12], the strain is applied in small increments, δ, (here we use δ = 2 x$10^{-4}$). Each incremental strain is followed by a minimization of the potential energy. The average shear stress for the quasi-static strain is also presented in Fig. 1 as a function of the accumulated strain. Instead of the relaxation of stress found for large step strains, the average stress under quasi-static strain approaches a constant non-zero value asymptotically in the strain [12]. The essential difference between the two strain protocols is that the step strain protocol allows for an unbounded affine strain energy from which to start the energy minimization. The quasi-static strain, in contrast, tightly constrains the amount of elastic energy that can be deposited into the solid prior to minimization. The quasi-static treatment clearly represents the more physically realistic scenario (i.e. one in which energy minimization occurs on time scales much quicker than that of the straining procedure) so why employ the step strain approach? There are two reasons. First, as discussed below, we are interested in generating non-affine displacements of arbitrary magnitude in order to explore the proposal that the rigidity of a solid is associated with particle constraint. The step strain protocol allows us to generate these non-affine strains of a wide range of magnitudes. The second virtue of the step strain is that it allows us to establish reversibility of strain in a simple and explicit manner, as we shall now discuss.

While the large strain behaviour of the step strain and quasi-static strain differ markedly, the small strain behaviour is very similar (see Fig. 1), even up to strains just beyond the yield value. This means that we can use the step strain calculations to study the role of irreversibility with regards the same nonlinear mechanical behaviour observed in the quasi-static approach over this range of strains.

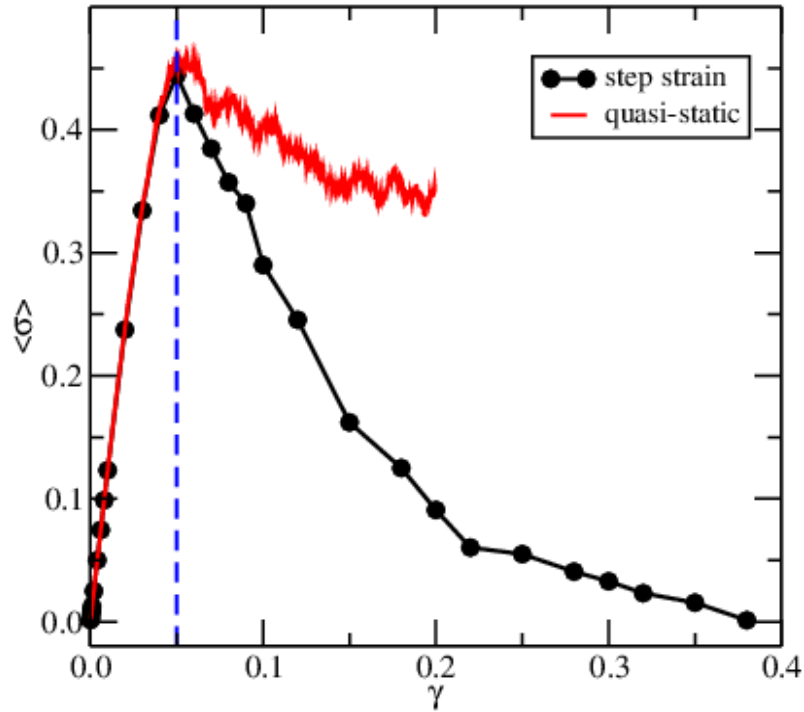

**Figure 1.** Plot of the average change in stress $<\sigma>$ vs $\gamma$ for the step strain and for the quasi-static strain. The yield strain $\gamma^*$ is indicated by the vertical dashed line.

A number of researchers have argued that nonlinear elastic response arises as a consequence of irreversible plastic events [12]. Our choice of a single step strain allows us to test directly for irreversibility as follows. Consider the following cyclic procedure: i) apply the affine strain, ii) minimize the potential energy, iii) reverse the affine strain, and iv) minimize the potential energy. If the final configuration satisfies the condition that no particle lies more than a reduced distance of 0.01 from its initial position, then we identify that process as reversible. In Fig. 2 (insert) we plot the fraction $f_{rev}(\gamma)$ of IS's that exhibit a reversibility for a shear strain of magnitude $\gamma$. At the yield strain $\gamma^* = 0.05$, we find that $f_{rev}(\gamma^*) = 0.32$, indicating that a third of the configurations reach the yield strain via a reversible deformation. It is possible, of course, that the observed mechanical nonlinearity is entirely due to the 68% of configurations that have undergone an irreversible rearrangement when $\gamma = \gamma^*$. To check whether this the case, we have plotted, in Fig. 2, the stress averaged over only those configurations that have exhibited a reversible strain at each value of γ using the protocol described above. We find that the reversible strains, while exhibiting a slightly larger yield stress, do so at essentially the same value of yield strain. We conclude that while irreversible reorganization at the yield strain does indeed describe the situation for the majority amorphous solids studied, irreversibility is not a *necessary* condition for yielding behaviour since reversible strains exhibit yield that is quantitatively similar to that generated with the inclusion of the irreversible strains. As is clear from Fig. 2, reversible strains are still observed even after the strain has increased past the peak in the stress.

We suggest that nonlinear elastic behaviour simply requires that the amplitude of particle displacement be large. While in crystalline materials, to achieve large enough displacements does indeed require irreversible reorganizations (i.e. defect motion), this is not in general the case in amorphous solids where the amplitude of nonaffine displacements is sufficient to achieve nonlinearity without an energy barrier being crossed. Non-affine displacements, i.e. the particle displacements resulting from the energy minimization of the configuration under an affine strain, are particularly important for the mechanical response of amorphous solids [13,14]. In Fig. 3 we plot the mean squared nonaffine displacement $<\Delta r^2>$ against γ. In the linear response regime, we expect $<\Delta r^2> \propto \gamma^2$ [14]. As shown in Fig. 3, we find, empirically, that this $\gamma^2$ relation applies well beyond the linear response regime. Just how large the displacements need to be to generate yield behaviour can be read directly off Fig. 3. At the field strain $\gamma^*$ we find that the nonaffine mean squared displacement is $<\Delta r^2> = 0.0478$. This value corresponds to an average displacement of ~0.2 of the small particle diameter. This is a substantial displacement, well beyond that typically associated with thermal motion in a stable solid. The demonstration here (see Fig. 3) that the magnitude of particle movement can be accomplished by a reversible strain underscores the remarkable magnitude of the non-affine motions accessed by amorphous solids.

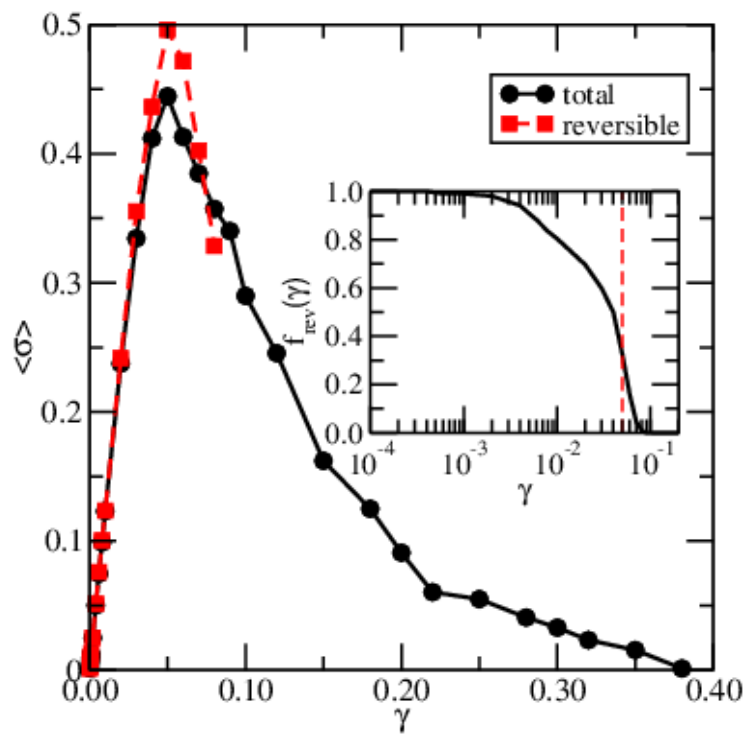

**Figure 2.** Plot of the shear stress <σ> vs applied strain γ calculated for the reversible strains (as explained in the text) (red squares) compared with the value from the all strains (black circles) Insert. The probability $f_{rev}(\gamma)$ of a strain being reversible as a function of the applied strain. The yield strain $\gamma^*$ is indicated by a vertical dashed line.

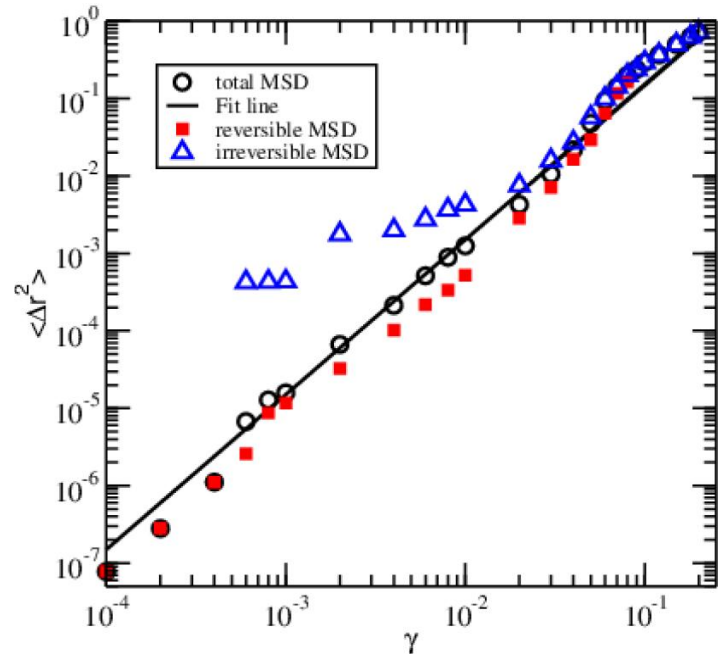

**Figure 3.** A plot of the non-affine $<\Delta r^2>$ vs $\gamma$. The values of $<\Delta r^2>$ for the reversible and irreversible strains are shown separately, along with the overall value. The straight line is a fit corresponding to $<\Delta r^2> \propto \gamma^2$.

## 4. On the Relaxation of Stress at T = 0 via Athermal Non-Affine Displacements.

A striking feature of the stress vs strain curves plotted in Fig. 1 is the steady decrease of $<\sigma>$ with increasing strain beyond the yield value. Previously [15], we have established that in a quiescent (i.e. unsheared) material, the loss of rigidity (as measured by the ratio of the shear and Born moduli $G_{eq}/G_\infty$) could be described by a function of only the magnitude of the thermal fluctuations in particle position $<\Delta r^2>$, independent of either the temperature or the observation time used to carry out the averages over the stress fluctuations. The conclusion drawn from this result was that rigidity is a consequence of the degree of configurational constraint experienced by a material and that the origin of the constraint – low temperature or short observation time – was not relevant. The stress relaxation reported here for the step strain calculations offer us an interesting opportunity to test this conclusion. We shall measure the overall rigidity of a sample subjected to the single step strain by the ratio $\langle\sigma\rangle/\gamma$. In the linear response regime, this ratio of stress over strain is simply the shear modulus $G_{eq}$. For $\gamma > \gamma^*$, $\langle\sigma\rangle/\gamma$ no longer equals the shear modulus [16] but it still describes the overall relationship between stress and strain. A characteristic of the step strain calculations is that they can generate large amplitude non-affine displacements, as shown in Fig. 3. In Fig. 4 we plot the reduced mechanical response function $\langle\sigma\rangle/(\gamma G_\infty)$ vs the non-affine mean squared displacement $\langle\Delta r^2\rangle$ from our T = 0 calculations along with the data for $G_{eq}/G_\infty$ from the quiescent liquid calculations in ref. 15. We find that the variation of the mechanical response $\langle\sigma\rangle/(\gamma G_\infty)$ with respects to the strain-induced $\langle\Delta r^2\rangle$ at T = 0 is very similar to the behaviour of the modulus $G_{eq}/G_\infty$ at non-zero temperatures due to thermally driven particle motion. This result provides significant support for the proposal [14] that rigidity is determined by the size of configuration space a system samples and that the details of how that space is sampled is not important.

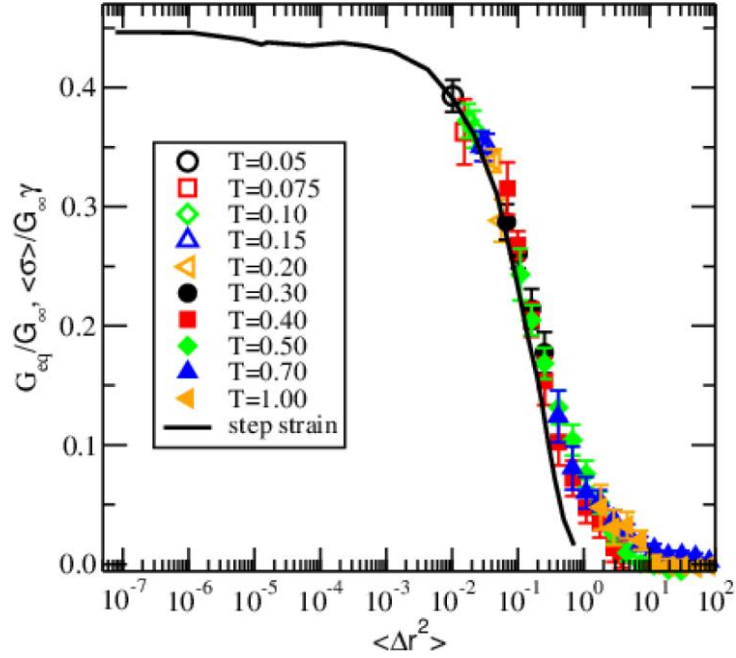

**Figure 4.** A plot of the reduced mechanical response $\dfrac{<\sigma>}{G_\infty \gamma}$ vs $<\Delta r^2>$ for the T = 0 solids

subjected to strain (solid line). Also plotted, the reduced shear modulus $G_{eq}/G_\infty$ from ref. 15, calculated over a range of temperatures and averaging times in quiescent liquids. The infinite frequency modulus $G_\infty = <\sigma_{affine}>/\gamma$, calculated using a strain of $10^{-4}$.

## 5. Conclusion

In this paper we have demonstrated that the nonlinear response of an amorphous solid with respect to a step-like shear strain can be completely accounted for by considering a measure, the mean squared non-affine displacement, of the extent of the configuration space sampled. The importance of the contribution of the collective strain in the nonlinear mechanical response of an amorphous solid has been previously noted [17]. Whereas previous studies have regarded plastic events as the essential drivers for these extended strains, here we have presented evidence that these irreversible reorganizations are not essential. We do not dispute that certain states (e.g. crystals) and certain shearing protocols (e.g. quasi-static shearing) constrain the magnitude of non-affine motion so that the only means of the system to achieve the necessary amplitude of particle movement is by irreversible rearrangements. Our argument is that it is useful to separate the process required to generate non-affine motion from the consequences of that non-affine motion once generated. This is a powerful result in that it asserts that the decrease in the shear modulus through nonlinear strain arises from the same fundamental cause as does the decrease in the modulus due to a temperature increase or an increase in observation time. Our results support the proposition that what matters is the magnitude of the non-affine displacements, not the means by which they are produced or constrained. We have presented evidence that the yielding of the amorphous solid is best described as a consequence of the shear-induced increase of the configuration space that is sampled . This picture includes, but is not limited to, the localized plastic events that have dominated much the discussion of yield behaviour. In our analysis, the significance of plastic reorganizations lies, not in their irreversibility, but, rather, in the fact that they represent an effective mechanism for increasing the size of the sampled configuration space (estimated here by the magnitude of $< \Delta r^2 >$) and the only possible mechanism in situations where non-affine motions are suppressed – either by high symmetry structure or by imposed constraints. A number of recent papers [18-21] have demonstrated that the rate of slow relaxation in glass forming liquids correlates strongly with the mean squared displacement, irrespective of whether particle motion is driven by thermal fluctuations [17,20] or an applied strain [21]. Clearly, a fundamental relationship must connect the origin of rigidity in configurational constraint, as discussed in this paper, and these correlations between displacements and dynamics, one that warrants further study.

**Acknowledgements**

This work was funded by a Discovery grant from the Australian Research Council with computer time provided by the National Computing Infrastructure.